\documentclass{aa}  

\usepackage{graphicx}
\usepackage{txfonts}
\usepackage{hyperref}
\usepackage{natbib}
\begin{document}

   \title{Ultra-pure digital sideband separation at sub-millimeter wavelengths}


   \author{R. Finger
          \inst{1,2}
          \and
          F.P. Mena\inst{2}
          \and
          A.Baryshev\inst{3,4}
          \and
          A.Khudchenko\inst{3}
          \and
          R.Rodriguez\inst{1}
          \and
          E.Huaracan\inst{1}
          \and
          A.Alvear\inst{1}
          \and
          J.Barkhof\inst{4}
          \and
          R.Hesper \inst{4}
          \and
          L.Bronfman\inst{1}
          }

   \institute{Astronomy Department, Universidad de Chile, Camino El Observatorio 1515, Santiago, Chile.\\
              \email{rfinger@u.uchile.cl}
         \and
             Electrical Engineering Department, Universidad de Chile, Avenue Tupper 2007, Santiago, Chile.
         \and
         	SRON Netherlands Institute for Space Research, Postbus 800, 9700 AV Groningen, The Netherlands.
         \and
         	NOVA/Kapteyn Astronomical Institute, University of Groningen, Postbus 800, 9700 AV Groningen, The Netherlands.
             }

   \date{}

 
  \abstract
   {Deep spectral-line surveys in the mm and sub-mm range can detect thousands of lines per band uncovering the rich chemistry of molecular clouds, star forming regions and circumstellar envelopes, among others objects. The ability to study the faintest features of spectroscopic observation is, nevertheless, limited by a number of factors. The most important are the source complexity (line density), limited spectral resolution and insufficient sideband (image) rejection (SRR). Dual Sideband (2SB) millimeter receivers separate upper and lower sideband rejecting the unwanted image by about 15 dB, but they are difficult to build and, until now, only feasible up to about 500 GHz (equivalent to ALMA Band 8). For example ALMA Bands 9 (602--720~GHz) and 10 (787--950~GHz) are currently DSB receivers.}
   {This article reports the implementation of an ALMA Band 9 2SB prototype receiver that makes use of a new technique called \textit{calibrated digital sideband separation}. The new method promises to ease the manufacturing of 2SB receivers, dramatically increase sideband rejection and allow 2SB instruments at the high frequencies currently covered only by Double Sideband (DSB) or bolometric detectors.}
   {We made use of a Field Programmable Gate Array (FPGA) and fast Analog to Digital Converters (ADCs) to measure and calibrate the receiver’s front end phase and amplitude imbalances to achieve sideband separation beyond the possibilities of purely analog receivers. The technique could in principle allow the operation of 2SB receivers even when only imbalanced front ends can be built, particularly at very high frequencies.}
   {This \textit{digital 2SB} receiver shows an average sideband rejection of 45.9~dB while small portions of the band drop below 40~dB. The performance is 27~dB (a factor of 500) better than the average performance of the proof-of-concept Band 9 purely-analog 2SB prototype receiver developed by SRON.}
   {We demonstrate that this technique has the potential of implementing 2SB receivers at frequencies where no such instruments exists, as well as improving the image rejection of current millimeter 2SB receivers to a level where sideband contamination is so low that would become negligible for any known astronomical source.}

   \keywords{Digital sideband separation --
                2SB --
                FPGA --
                ALMA Band 9 --
                Molecular Spectroscopy --
                Confusion limit.
               }

   \maketitle
%

\section{Introduction}

Astronomy is one of the main drivers of high-frequency radio technology. Astronomical radio receivers have the most demanding specifications since they require very low noise, high stability and extremely large bandwidths. The ALMA observatory is the pinnacle example of such a technology driver. It has implemented 66 radio telescopes, each one with 10 observing bands covering frequencies from 35 to 950~GHz, fractional bandwidths ranging from 20 to 35\%, and receiver noise temperatures just a few times above the absolute quantum limit \citep{wootten2009}. 

Modern single dish telescopes and arrays commonly have large collecting areas which, together with low receiver noise and high/dry locations, produce exceptional sensitivities allowing scientists to probe deeper than ever in the faintest spectroscopic features of astronomical sources. In fact, deep spectral line millimeter surveys can detect thousands of lines per band uncovering the rich chemistry of molecular clouds, star forming regions and circumstellar envelopes, among others objects \citep{cernicharo2013,nimesh2011}. 
For example, in 2010, \citet{tercero2010}, reported the detection of more than 14400 spectral lines in Orion-KL, observing the 80$-$115.5~GHz, 130$-$178~GHz, and 197$-$281~GHz bands. Lines intensities ranged from around 10~mK to almost 100~K, i.e. an intensity ratio of almost 10.000 or 40~dB.

\begin{figure}[ht!]
\centering
\includegraphics[width=\hsize]{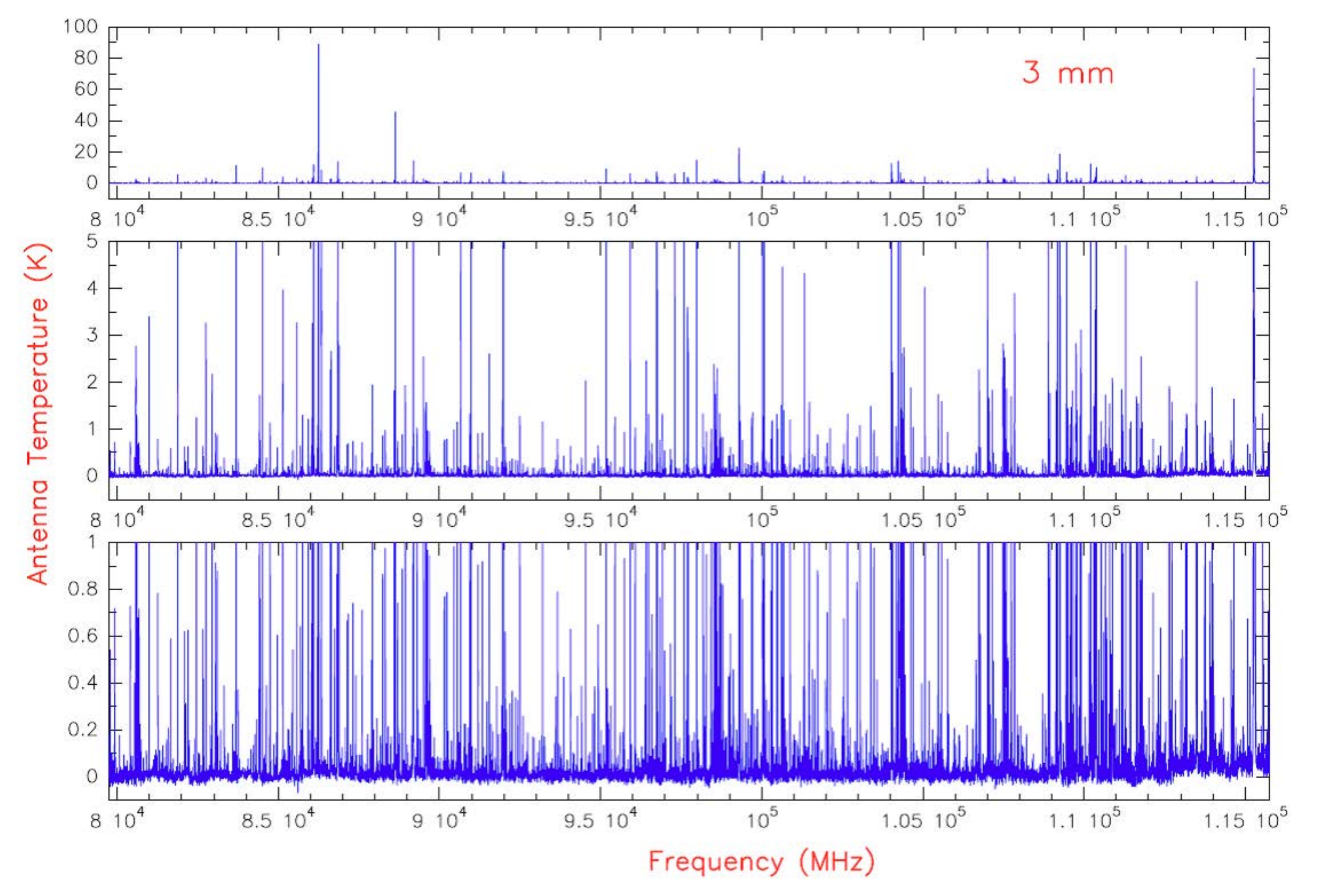}
\caption{A confusion limited spectroscopic survey of Orion-KL performed by Tercero, et Al. with the IRAM 30~m telescope. Taken from \citet{tercero2010}. From top to bottom the panels show the same spectrum at 3 different vertical scales. Lines intensities range from $\sim$10~mK to almost 100~K.}
\label{fig:fig1}
\end{figure}

The ability to study the faintest features of spectroscopic observation is, nevertheless, limited by a number of factors. The most important are the source complexity (line density), limited spectral resolution, and insufficient sideband (image) rejection. These conditions set what is known as the confusion limit, i.e. the level below which the merging of a huge number of weak lines makes impossible any further discrimination.

For observing complex sources, as the one presented in figure \ref{fig:fig1}, it is desirable to have single sideband (SSB) or, ideally, 2SB receivers, with SRR in excess of 30~dB and preferably above 40~dB. Despite the clear importance of having high sideband rejection for deep spectroscopic surveys, the specifications set by ALMA for its bands 1 to 8 is only 7~dB across the band, and 10 dB for the 80\% of the band \citep{alma_specs2001,ALMA_CPB_2002}. The reason for requiring what might seems a low performance was based on the extreme difficulty of achieving higher suppressions, due to the need of keeping a flat phase and amplitude response, across wide bands at high frequencies.

Until recently 2SB receivers were only available at relatively low frequencies, below 300~GHz. ALMA pushed that limit, requiring its Band 8 (500~GHz) to be sideband separating. Nevertheless, ALMA receivers above 600~GHz (Bands 9 and 10) are DSB and, therefore, have limited capabilities to disentangle sources with a high line density, and a reduced sensitivity due to the noise contribution from the sideband. 

A new technique that promises to ease 2SB receivers manufacturing as well as dramatically increase the sideband rejection performance has been proposed by \citep{morgan2010} and implemented by \citep{finger2013}. In this approach the IF hybrid is removed, replaced by two Analog-to-Digital Converters (ADC), and implemented in-digital within a Field Programmable Gate Array (FPGA). The digital implementation of the IF hybrid allows to calibrate-out imbalances of the analog components to achieve very high sideband rejection. The calibration of RF imbalances eases the requirements on phase flatness for RF hybrids and gain flatness for mixers and amplifiers, allowing the operation of 2SB receivers at frequencies and bandwidths where purely analog technics are impractical.

Digital receivers of this sort have been demonstrated in laboratory up to 100~GHz \citep{rodriguez2014,rodriguez2015_thesis} showing a SRR in excess of 45 dB. This corresponds to 20 to 30~dB (a factor 100 to 1000) above state-of-the-art performance of analog 2SB receivers.

This article reports the integration of an FPGA digital sideband separating spectrometer with an ALMA Band 9 2SB prototype receiver to demonstrate, for the first time, calibrated digital sideband separation at sub-millimeter wavelengths.


\section{Measurement setup and procedure}

\begin{figure}[ht!]
\includegraphics[width=\hsize]{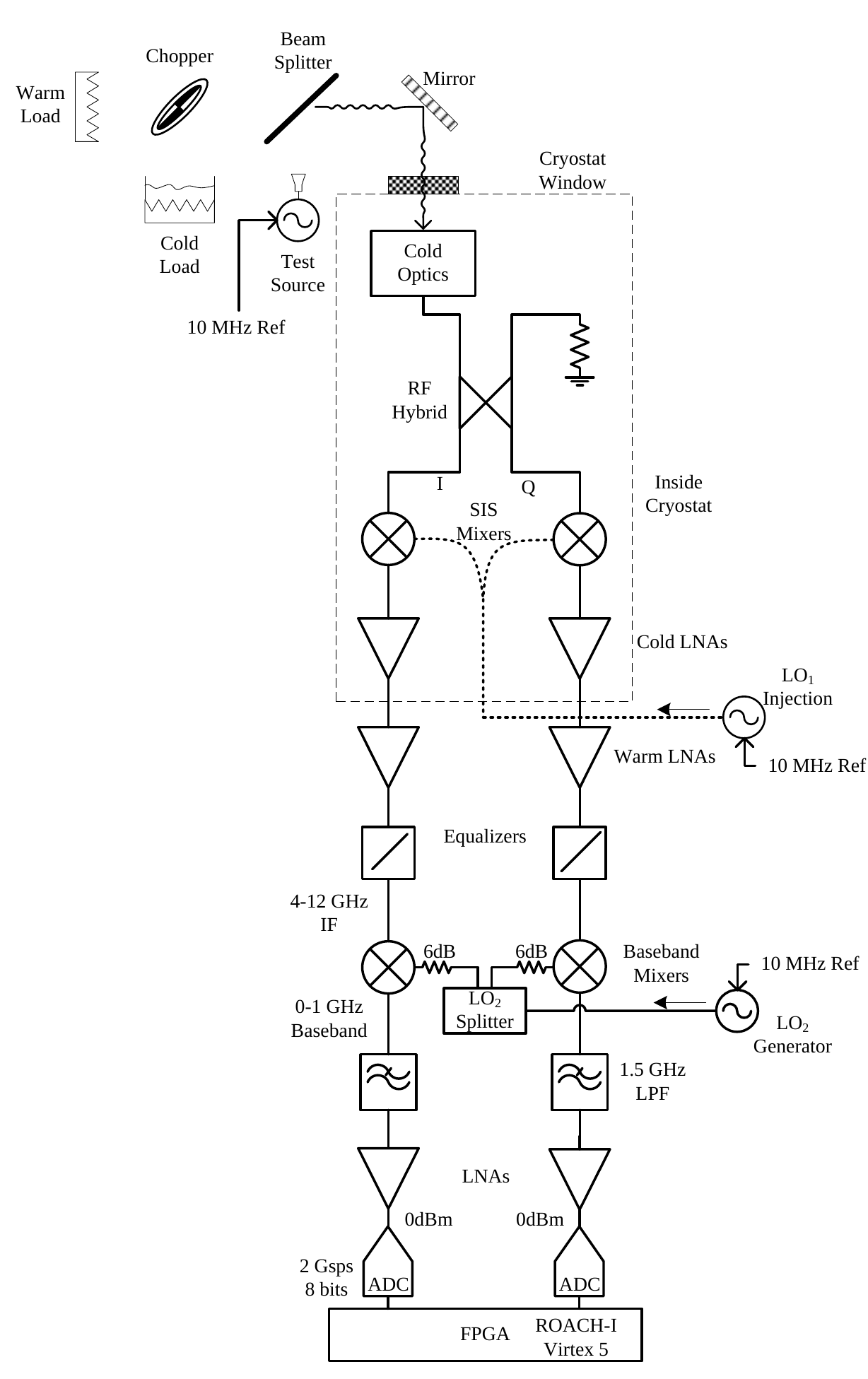}
\caption{Test setup block diagram: The front end of the receiver features an RF hybrid which splits the incoming signal into the I and Q channels. A first down-conversion to a 4--12~GHz IF band is performed using a pair of superconducting (SIS) mixers driven by an in-phase local oscillator (LO$_{1}$). Cold and warm LNAs are used to boost the signal and then equalizers compensate the amplifiers gain slope. A second down-conversion to a 0--1~GHz baseband is performed using a pair of balanced mixers driven by an in-phase second local oscillator (LO$_{2}$). Finally anti-aliasing filters and a last amplification stage are applied to reach optimal signal condition for the digitizers. The test source and the local oscillators are locked to the same 10 MHz reference to prevent frequency drift of the test tone during calibration. Recombination of the I and Q channels is done in the FPGA to produce the USB and LSB spectra.}
\label{fig:fig2}
\end{figure}

A digital sideband separating receiver was configured using as front-end an ALMA Band 9 2SB prototype receiver that uses AlN junctions as mixing elements \citep{khudchenko2011}. The only modification respect to the original receiver was the removal of the IF hybrid to access the receiver’s I and Q channels independently. The experimental setup, shown in figure \ref{fig:fig2}, allows the input signal to be injected into the receiver with a cold and ambient background using a beam-splitter and a chopper.

The IF outputs of the I and Q SIS mixers are amplified, equalized and then mixed again using a pair of DSB mixers and a second LO (LO$_{2}$) which selects any 1~GHz segments within the 4--12~GHz range to be down-convert to the 0--1~GHz baseband. Two 1.5~GHz low pass filters are used after the second downconversion to reject out-of-band noise and spurious, as well as to prevent LO$_{2}$ leakage getting into the ADCs. Since the second downconversion is DSB, noise and spurious signal coming from warm LNAs and the LO$_{2}$ will not be suppressed. Sufficient gain in the cold LNAs should render the IF-chain noise contribution negligible.

The measurement procedure requires an initial calibration run which measures the amplitude and phase imbalances of the entire analog front end. These imbalances are then compensated for each spectral channel in the FPGA. The balanced I and Q mixer outputs are then recombined in a digital IF hybrid to produce two (LSB and USB) 2048 channels spectra with very high sideband rejection.
After calibration a test tone is swept across the RF band and measurements are taken with ambient and cold backgrounds across the IF. The data is then used to calculate the sideband rejection using the method described in \citep{alma_memo_357} which allows an SRR measurement independent of the test tone power. Details of this technique are reported in \citep{finger2013} and \citep{rodriguez2014,spie_2014}. Typical spectra measured using the prototype ALMA Band-9 receiver, are presented in figure \ref{fig:fig3}.

\section{Factors limiting sideband rejection}

\begin{figure}[ht!]
\includegraphics[width=\hsize]{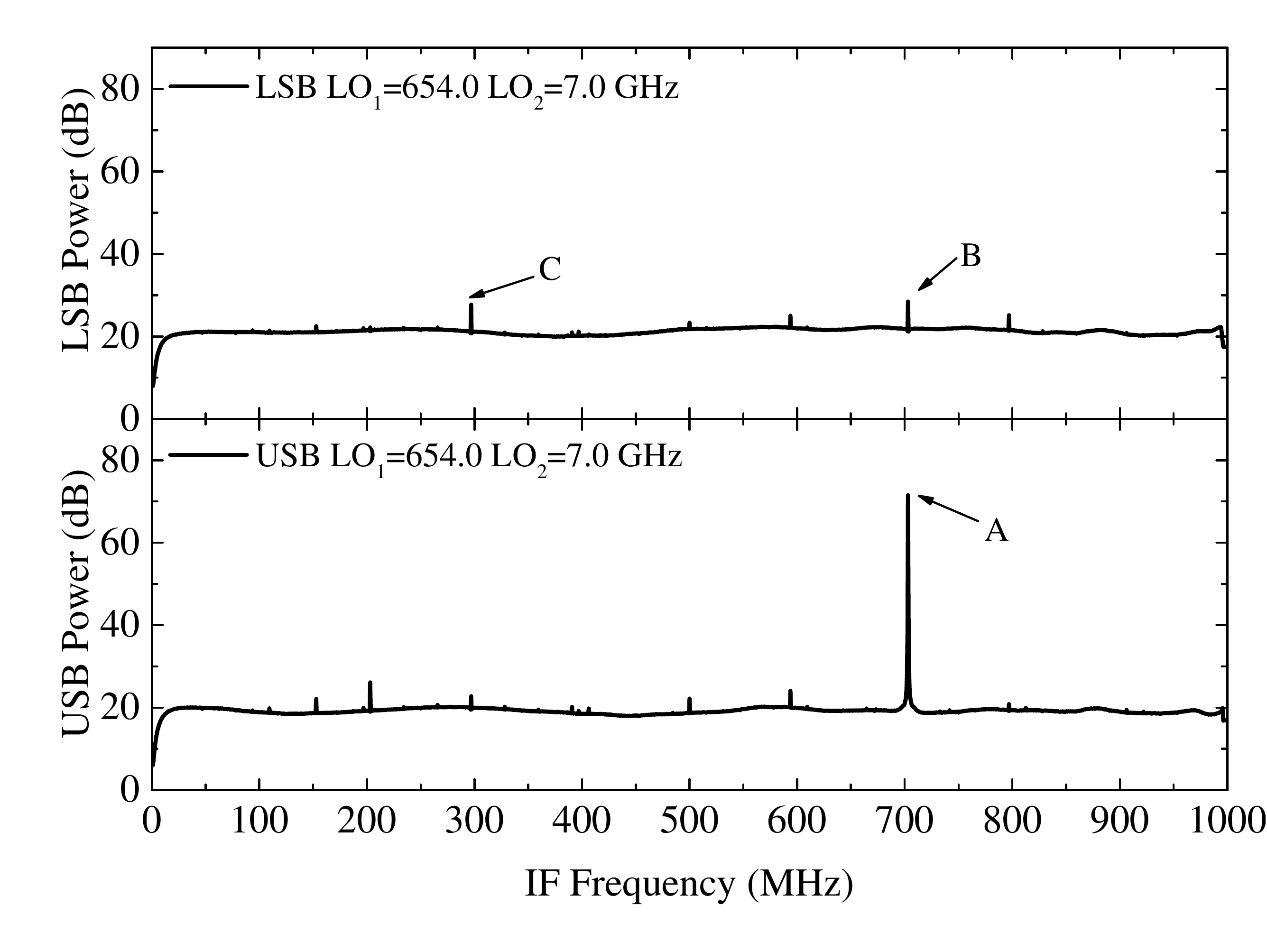}
\caption{Typical spectra for the calibrated receiver. Signals A, B and C correspond to the test tone in the pass-band, the tone in the rejected-sideband, and the highest spurious (intermodulation product) in the rejected-sideband, respectively.}
\label{fig:fig3}
\end{figure}

There are two main effects causing the reduction in digital sideband rejection in this setup. First, the poor isolation between the RF and LO ports in the baseband mixers and within outputs in the LO$_{2}$ splitter. Second, the intermodulation products of LO$_{1}$ and test source harmonics. Following figure \ref{fig:fig2}, the first issue is that the RF signal entering the baseband mixer I leaks to its LO port and is passed to channel Q through the LO$_{2}$ splitter. The same happens in the opposite direction, i.e from channels Q to I. This leaked signal has the same frequency than the main RF signal on each branch, but having a different phase, it will not add up in counter-phase in the calibrated digital IF hybrid. Consequently, it will not be cancelled out. This unrejected signal is detected as a small tone in the order of -40~dBc at the same channel, in the rejected sideband (tone B in fig. \ref{fig:fig3})). Two 6~dB fixed attenuators were placed in the splitter outputs to increase isolation by an additional 12~dB. In this configuration the spurious tone is attenuated below the test setup noise floor, as seen in figure \ref{fig:fig3}.

In addition to the unrejected image tone B, another spurious signal, which is caused by intermodulation of the RF test source and the first LO harmonics, can be seen (tone C in fig. 3). They appear because the LO and RF sources are based on $\times$3 and $\times$9 multipliers and harmonics up to -20~dBc can be generated. In combination with the high nonlinearity of the SIS-mixers parasitic signals are produced. Even though these spurious signals are not limiting the suppression of the main test tone, it does limit the spurious free dynamic range of this receiver to about -45~dB. As we will explain below, it does also produce false reading of SRR when it falls into the same channel where the test tone is tuned.

Even with a perfectly clean LO, and fully isolated I and Q channels, there are others, more fundamental factors, limiting the sideband rejection than can be achieved by this technique. The most critical is the Spurious Free Dynamic Range (SFDR) of the ADC chips. For the hardware used in this experiment the average SFDR is close to 55~dB, so it is not the limiting factor here, but for lower frequency receiver, where low harmonic content and high isolation can be achieved, the ADC SFDR has been reported as the limiting factor for a usable sideband rejection \citep{finger2013}.

Finally, numeric error due to the limited number of bits available for data representation within the FPGA and the calibration error due to the limited spectral resolution are possible limiting factors for SRR. In one hand, we have and 18+18 bits (real and imaginary) data type in the FFT, which limits the sideband rejection to about 100~dB, so it is not a practical limitation. On the other hand the calibration error depends on the gain and phase slopes to be compensated within the 1~GHz baseband. If we assume a receiver gain variation less than 3dB$/$GHz and a phase ripple less than 90$^\circ/$GHz the limit to the SRR imposed by our 1024 points calibration is in the order of 70~dB.
Both limits are considerable above the SFDR of the fast ADCs needed for this application, so the practical limit for usable SRR is imposed by ADC SFDR performance.


Summarizing, SRR could be improved by redesigning of the LO$_{1}$ and RF test-source to reduce harmonic content and also modifying the LO2 splitting scheme for better I-Q isolation. I-Q isolation can be improved by adding isolators or amplifiers after the LO$_{2}$ splitter. LO$_{1}$ and RF-source spurious content could be reduced by adding waveguide filters after each multiplication stage. If both modifications are applied, we could improve SRR by an additional 5--10~dB where we will hit the 55~dB limit imposed by the ADCs SFDR.

\section{FPGA firmware}

\begin{figure}[ht!]
\includegraphics[width=\hsize]{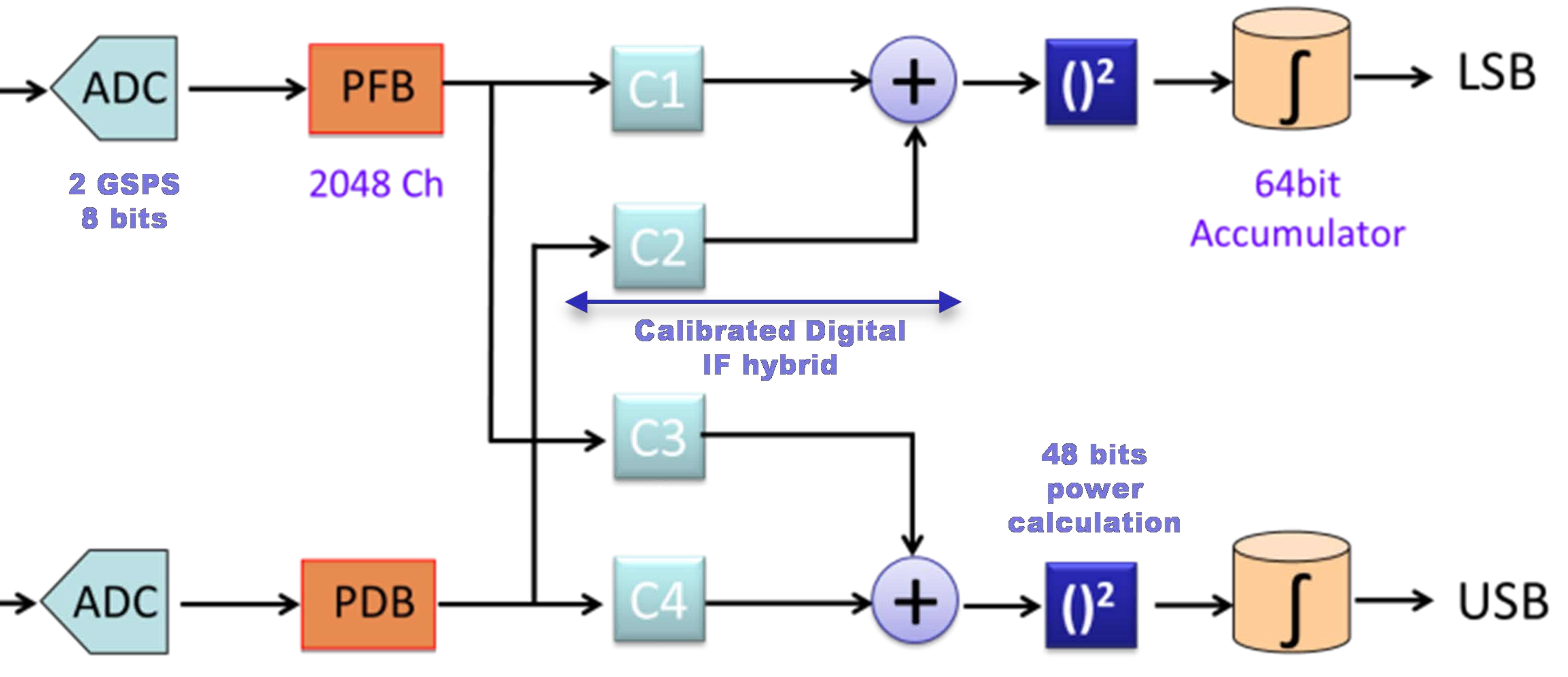}
\caption{FPGA firmware top-level block diagram. From left to right main subsystems are: ADC hardware interface, Polyphase Filter Bank (PFB),Complex vector multipliers (C1-C4), Complex Adders (+), Power blocks ( )$^2$, and vector accumulators.}
\label{fig:fig4}
\end{figure}

\begin{figure}[ht!]
\includegraphics[width=\hsize]{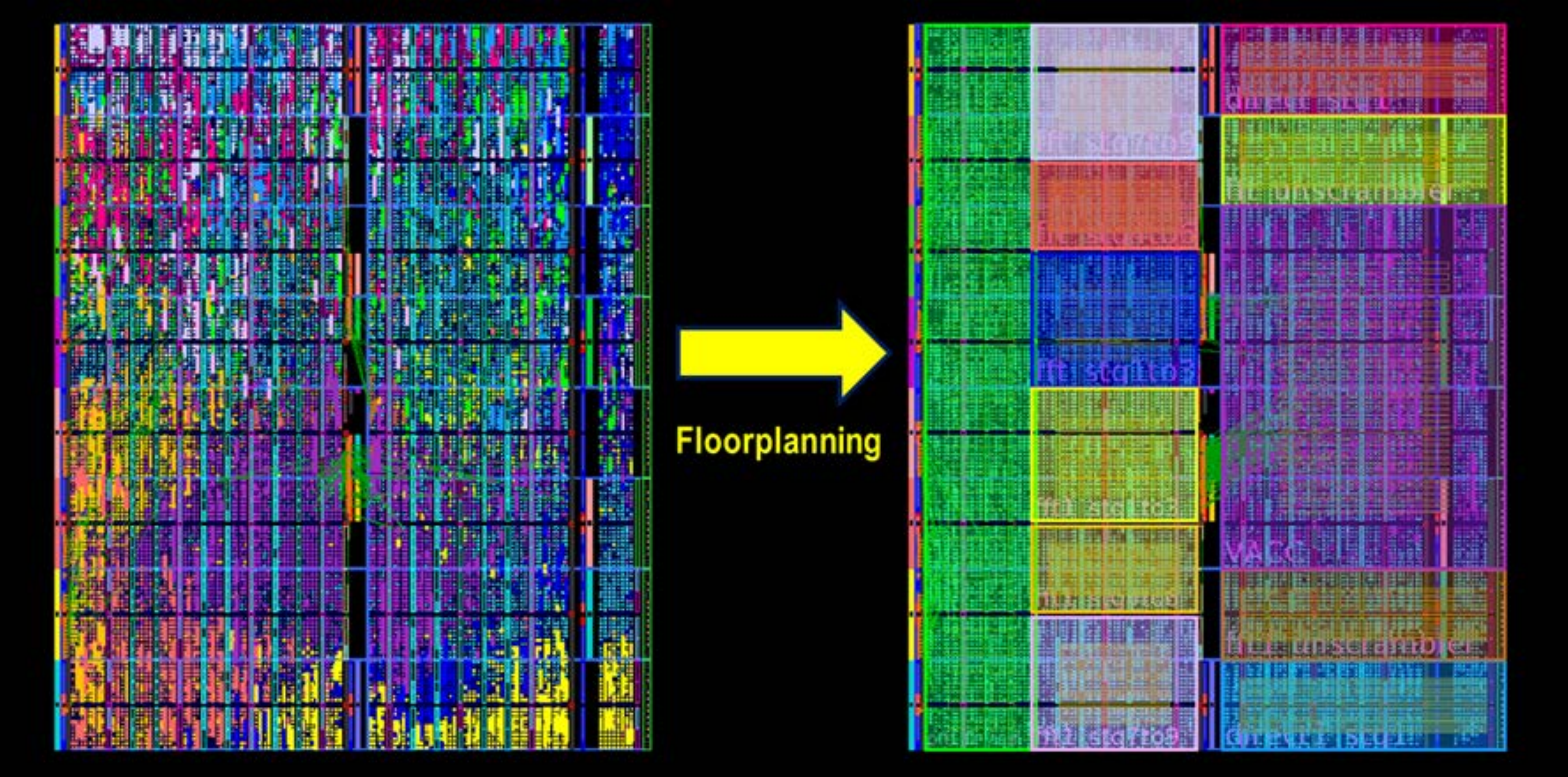}
\caption{Logic resources allocation in the FPGA (Virtex 5) chip surface before and after \textit{floorplanning}. Each color represents a processing subsystem depicted in figure \ref{fig:fig4}. Automatic place and routing algorithm implements the processing subsystems using logic resources scattered across the chip. \textit{Floorplaning} optimizes resource location to minimize delay and delay-difference through the parallel processing pipeline.}
\label{fig:fig5}
\end{figure}

The ROACH-I FPGA open platform \citep{casper} was used to implement the sideband-separating spectrometer. The FPGA firmware used here, and presented schematically in figure \ref{fig:fig4}, was similar in function to the one described in \citep{finger2013} and \citep{rodriguez2014,spie_2014}. However, for this work, it features optimized versions of the calibration and sideband separating spectrometers. These enhanced spectrometers were implemented using a technique called \textit{floorplanning}, which involves human-aided physical disposition of logical resources over the FPGA chip area, to minimize delays and equalize data paths lengths. Using this technique, faster clock speeds of those reachable by the use of automatic compilation tools can be reached.
The software used to perform the \textit{floorplanning} was Xilinx ISE, PlanAhead. The \textit{floorplanning} strategy consisted in defining areas in the chip to target the compilation of the spectrometers different subsystems, and localize those areas in a way consistent with the data flow structure. This not only reduces data path lengths (and therefore delays) within subsystems in the chip, but also the delay-difference between parallel processing branches. The optimized design (see figure \ref{fig:fig5}) doubled the clock frequency achievable by the automatic compilation tools, reaching an instantaneous bandwidth of 1~GHz per sideband \citep{alvear2013}.

\section{Results}

\begin{figure}[ht!]
\includegraphics[width=\hsize]{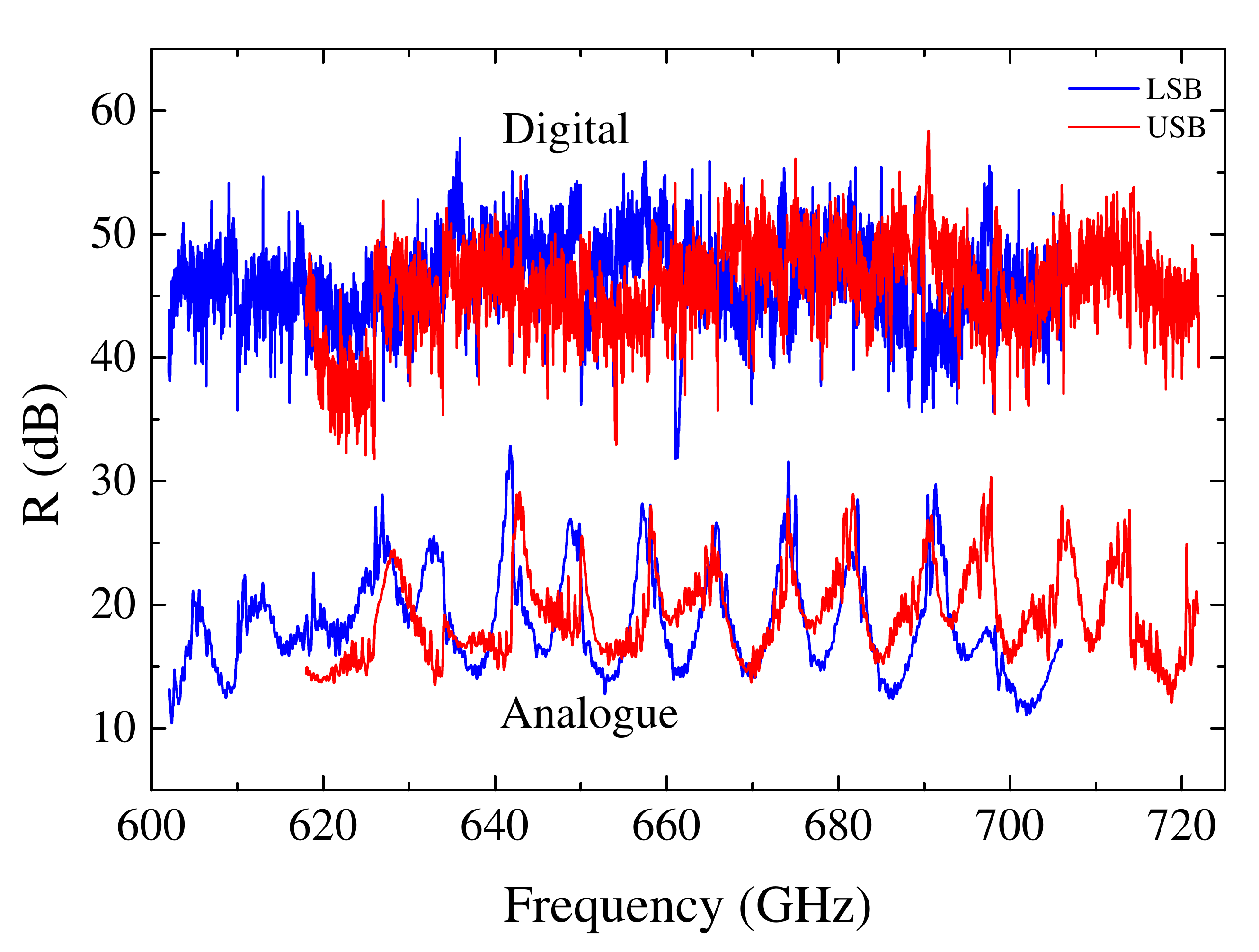}
\caption{Calibrated digital sideband rejection ratio. LO$_{1}$ was swept from 614 to 710~GHz with 8 GHz steps. LO$_{2}$ was swept from 4 to 11~GHz for each LO$_{1}$ setting. Analogue sideband rejection performance \citep{khudchenko2011} is shown for comparison.}
\label{fig:fig6}
\end{figure}

Figure \ref{fig:fig6} shows the SRR for the entire ALMA Band-9 RF band using the technique presented above. LO$_{1}$ was swept from 614 to 710~GHz with 8~GHz steps, while LO$_{2}$ was swept from 4 to 11~GHz for each LO$_{1}$ setting. Analog sideband rejection performance \citep{khudchenko2011} is shown for comparison.

\begin{figure}[ht!]
\includegraphics[width=\hsize]{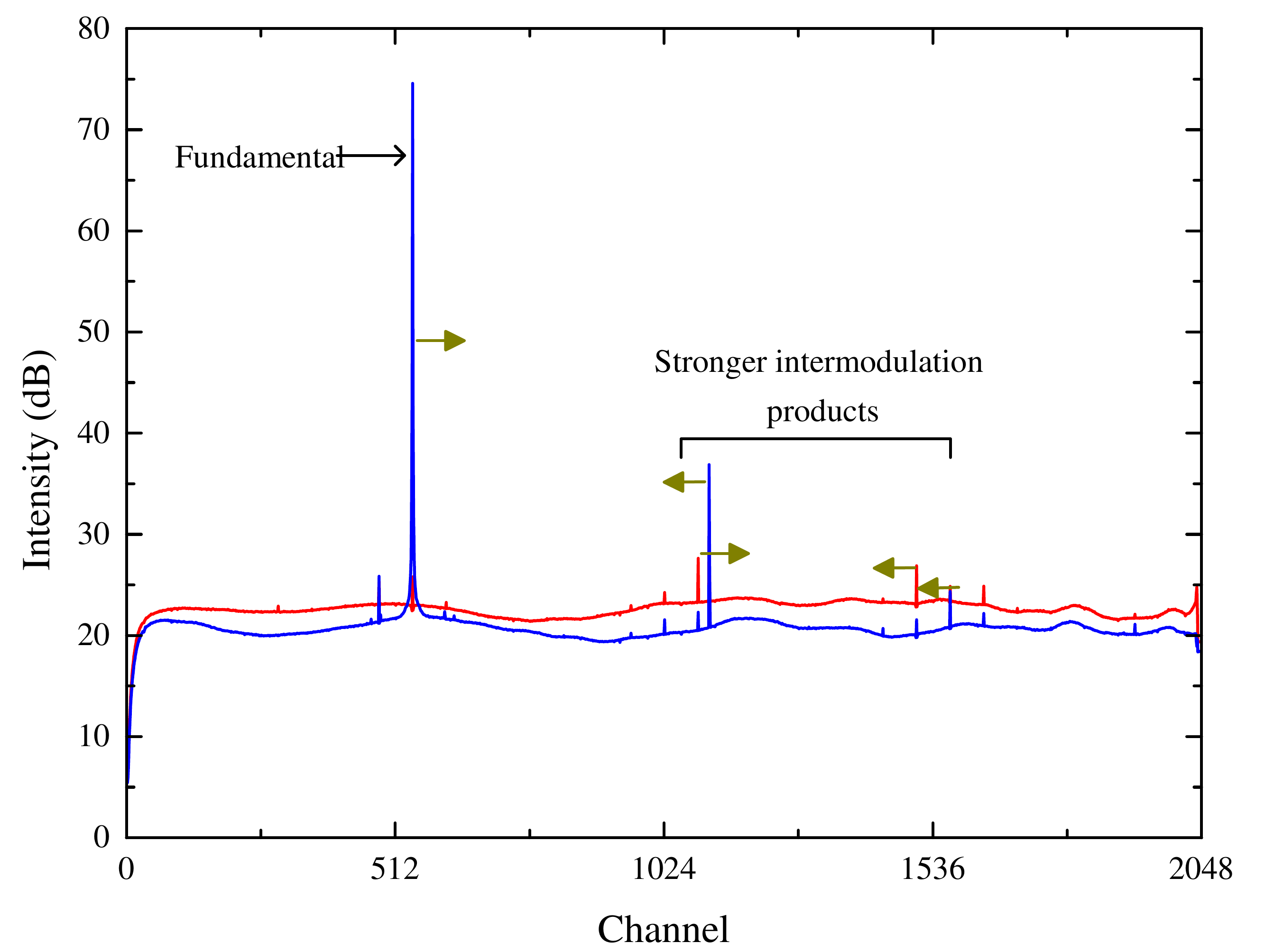}
\caption{A case of strong spurious contamination for LO$_{1}$=686. Arrows indicate the direction of movement of each spurious signal while sweeping up the fundamental tone.}
\label{fig:fig7}
\end{figure}

The average performance of the digital implementation is around 45~dB while small portions of the band drop below 40 dB. The high dispersion in the data as well as the individual points that drop below the average are explained caused by strong spurious coming from LO$_{1}$ and test source harmonics as seen in figure \ref{fig:fig7}. These intermodulation products overlap the test tone fundamental frequency in the rejected band, generating false readings of the sideband rejection. These several spurious signals move in both directions when the test tone is sweep across the IF, so the overlaps with the rejected tone occur at random-looking frequencies.

The amplitude of such intermodulation product is highly dependent on frequency; consequently for some LO settings the dispersion and dropping points are larger than for others.

Finally, the more consistent drop of SRR measured on LSB between 620 and 630~GHz is explained by a reduction in signal to noise ratio due to a drop in the test tone power. This produces the fundamental tone to decrease in amplitude while the rejected tone was already below noise floor. It is important to mention that an integration of 134 milliseconds was used in our setup. More integration time would help SRR measurements in this part of the band.

\section{On the calibration stability}

Thermal stability of the calibration has been demonstrated in \citep{finger2013} by thermal cycling the IF plate by 10~K for three times with no measurable degradation of SRR. In this work, the RF part of the receiver is inside a cryostat with very good (0.1~K) thermal stability so we do not see a need of recalibrating the receiver due to thermal stability. Stability of the calibration after SIS mixers defluxing can be partially estimated from the work of \citep{Bout2008}, where SIS junctions were subjected to 40 defluxing cycles with very minor changes in conversion characteristics, warrantying a stability better than 0.1~dB for AlN junctions. This stability ensures an SSR above 40~dB without a new calibration. SIS phase delay stability after defluxing has not been particularly evaluated in published works, but ALMA and other observatories, which use analogue 2SB SIS receivers, have not reported significant SRR degradation to our knowledge, which suggest that phase delay of SIS junction do not change by important amounts. Therefore, we considered that recalibration will not be needed frequently and probably recalibration after each cryostat thermal cycle would be sufficient.

\section{Conclusions}

An ALMA Band-9 2SB prototype mixer was integrated with a digital sideband separating spectrometer based on FPGA. After calibration, full RF/IF band measurements were performed achieving an average sideband rejection of 45.9~dB and above 40~dB for 93\% of the band. Although some frequency points dropped below 35~dB, this is explained as limitations of the test setup and LO$_{1}$ source purity. Therefore, the digital receiver performance exceeded the analog receiver SRR by 27~dB on average.

The achieved SRR is high enough so that sideband contamination becomes negligible for any known astronomical application.

The ability of calibrating this submillimeter receiver achieving very high sideband rejection suggests that higher frequency 2SB receivers might be implemented using this technique, reaching image rejection levels otherwise impossible by purely analog means.
	
The measurements have also shown that to observe complex sources with line intensity ratios in excess of 30~dB, LO cleanness might be an issue at submillimeter wavelengths.

Even though the instantaneous bandwidth of this digital receiver is 1~GHz per sideband, scaling the bandwidth up has no conceptual limitations and involves only stacking several digital boards.

\begin{acknowledgements}
This work has been possible thanks to the support of CONICYT through the funds CATA-Basal PFB06 and FONDECYT 11140428 and 1121051. We also thank Xilinx Inc. for the donation of FPGA chips and software licenses as well as the whole CASPER community for its invaluable support.
\end{acknowledgements}

\bibliographystyle{aa}
\bibliography{references}
%
%
%
%
%
%
%
%
%

\end{document}